\documentstyle[preprint,psfig,aps,floats,tighten]{revtex} 

\begin{document}

\preprint{Fermilab-Pub-97/172-E}
\title{Measurement of the Top Quark Mass Using Dilepton Events}

\author{
\centerline{The D\O\ Collaboration\thanks{Authors listed on the following page.
            \hfill\break 
            Submitted to Physical Review Letters.}}
}

\address{
\centerline{Fermi National Accelerator Laboratory, Batavia, Illinois 60510}
}

\date{June 9, 1997}

\maketitle

\begin{abstract}
The D\O\ collaboration has performed a measurement of the top quark mass
 $m_t$ based on six candidate events
for the process $t\overline{t} \rightarrow bW^+ \overline{b}W^-$, where
the $W$ bosons
decay to $e\nu$ or $\mu\nu$.  This sample was collected during
an exposure of the D\O\ detector to an integrated luminosity of 
 125~$\mbox{ pb}^{-1}$ of $\sqrt{s} = 1.8$ TeV
$p\overline{p}$ collisions.
We obtain $m_t = 168.4 \pm 12.3 \mbox{ (stat)} \pm  
3.7 \mbox{ (sys)}$
GeV/{\it c}$^2$, consistent with the measurement obtained using
single-lepton events.  Combination of the single-lepton and dilepton results yields
$m_t = 172.0 \pm 7.5$ GeV/{\it c}$^2$.
\end{abstract}
\pacs{PACS numbers: 14.65.Ha, 13.85.Qk, 13.85.Ni}

\vskip 1cm

\begin{center}
B.~Abbott,$^{28}$                                                             
M.~Abolins,$^{25}$                                                            
B.S.~Acharya,$^{43}$                                                          
I.~Adam,$^{12}$                                                               
D.L.~Adams,$^{37}$                                                            
M.~Adams,$^{17}$                                                              
S.~Ahn,$^{14}$                                                                
H.~Aihara,$^{22}$                                                             
G.A.~Alves,$^{10}$                                                            
E.~Amidi,$^{29}$
% ...
\\                                                                            
\end{center}

\vskip 0.50cm                                                                 
\normalsize
\centerline{(D\O\ Collaboration)}                                             
\vfill\eject                                                                 
\small
\it                                                                           
\centerline{$^{1}$Universidad de los Andes, Bogot\'{a}, Colombia}             
\centerline{$^{2}$University of Arizona, Tucson, Arizona 85721}               
\centerline{$^{3}$Boston University, Boston, Massachusetts 02215}             
\centerline{$^{4}$Brookhaven National Laboratory, Upton, New York 11973}      
\centerline{$^{5}$Brown University, Providence, Rhode Island 02912}           
\centerline{$^{6}$Universidad de Buenos Aires, Buenos Aires, Argentina}       
\centerline{$^{7}$University of California, Davis, California 95616}          
\centerline{$^{8}$University of California, Irvine, California 92697}         
\centerline{$^{9}$University of California, Riverside, California 92521}      
\centerline{$^{10}$LAFEX, Centro Brasileiro de Pesquisas F{\'\i}sicas,        
                  Rio de Janeiro, Brazil}                                     
\centerline{$^{11}$CINVESTAV, Mexico City, Mexico}                            
\centerline{$^{12}$Columbia University, New York, New York 10027}             
\centerline{$^{13}$Delhi University, Delhi, India 110007}                     
\centerline{$^{14}$Fermi National Accelerator Laboratory, Batavia,            
                   Illinois 60510}                                            
\centerline{$^{15}$Florida State University, Tallahassee, Florida 32306}      
\centerline{$^{16}$University of Hawaii, Honolulu, Hawaii 96822}              
\centerline{$^{17}$University of Illinois at Chicago, Chicago,                
                   Illinois 60607}                                            
\centerline{$^{18}$Indiana University, Bloomington, Indiana 47405}            
\centerline{$^{19}$Iowa State University, Ames, Iowa 50011}                   
\centerline{$^{20}$Korea University, Seoul, Korea}                            
\centerline{$^{21}$Kyungsung University, Pusan, Korea}                        
\centerline{$^{22}$Lawrence Berkeley National Laboratory and University of    
                   California, Berkeley, California 94720}                    
\centerline{$^{23}$University of Maryland, College Park, Maryland 20742}      
\centerline{$^{24}$University of Michigan, Ann Arbor, Michigan 48109}         
\centerline{$^{25}$Michigan State University, East Lansing, Michigan 48824}   
\centerline{$^{26}$Moscow State University, Moscow, Russia}                   
\centerline{$^{27}$University of Nebraska, Lincoln, Nebraska 68588}           
\centerline{$^{28}$New York University, New York, New York 10003}             
\centerline{$^{29}$Northeastern University, Boston, Massachusetts 02115}      
\centerline{$^{30}$Northern Illinois University, DeKalb, Illinois 60115}      
\centerline{$^{31}$Northwestern University, Evanston, Illinois 60208}         
\centerline{$^{32}$University of Notre Dame, Notre Dame, Indiana 46556}       
\centerline{$^{33}$University of Oklahoma, Norman, Oklahoma 73019}            
\centerline{$^{34}$University of Panjab, Chandigarh 16-00-14, India}          
\centerline{$^{35}$Institute for High Energy Physics, 142-284 Protvino,       
                   Russia}                                                    
\centerline{$^{36}$Purdue University, West Lafayette, Indiana 47907}          
\centerline{$^{37}$Rice University, Houston, Texas 77005}                     
\centerline{$^{38}$Universidade Estadual do Rio de Janeiro, Brazil}           
\centerline{$^{39}$University of Rochester, Rochester, New York 14627}        
\centerline{$^{40}$CEA, DAPNIA/Service de Physique des Particules,            
                   CE-SACLAY, Gif-sur-Yvette, France}                         
\centerline{$^{41}$Seoul National University, Seoul, Korea}                   
\centerline{$^{42}$State University of New York, Stony Brook,                 
                   New York 11794}                                            
\centerline{$^{43}$Tata Institute of Fundamental Research,                    
                   Colaba, Mumbai 400005, India}                              
\centerline{$^{44}$University of Texas, Arlington, Texas 76019}               
\centerline{$^{45}$Texas A\&M University, College Station, Texas 77843}       
\normalsize

\vfill\eject
  The pair production of top quarks has been observed in $p\overline{p}$
collisions at $\sqrt{\rm{s}}$ = $1.8$ TeV by the CDF and D\O\ collaborations
\cite{top_observation}.  Since the time of observation,
the integrated luminosity has more than doubled (to 125 pb$^{-1}$) and
the D\O\ experiment has substantially improved its techniques for measurement
of the top
quark mass $m_t$.  We previously reported a measurement of $m_t$ using 
events in which one top quark decayed semileptonically and the other decayed
hadronically (the ``$\ell$ + jets'' mode, where $\ell = e$ or $\mu$), giving
$m_t = 173.3 \pm 5.6 \mbox{ (stat)} \pm 6.2 \mbox{ (sys) GeV/}c^2$
\cite{ljets_mass_PRL}.
 This letter reports a first
measurement of $m_t$ using events consistent with the
$t\overline{t} \rightarrow  bW^+\overline{b}W^- \rightarrow
b\ell^+ \nu\overline{b}\ell^- \overline{\nu}$ (``dilepton'') 
hypothesis.  This independent measurement is important as a direct test of
the hypothesis 
that the excess of events over background in both the $\ell$ + jets and
 dilepton channels is due to $t\overline{t}$ production.

  The events used in this analysis were recorded by the D\O\ detector 
\cite{D0_NIM_paper}, which consists of a nonmagnetic tracking system,  
including a
transition radiation detector (TRD), 
surrounded by a hermetic liquid argon/uranium calorimeter, segmented in depth
into several electromagnetic (EM) and hadronic layers, and an outer toroidal 
muon spectrometer.  Electrons are identified using a likelihood method
based on the EM shower shape, track ionization, spatial match of the track with
the EM shower, and TRD response.  Muons are
required to have reconstructed tracks in the central tracking chamber and
in at least one of the spectrometer layers outside
of the toroid, and to
have energy deposition in the calorimeter consistent with the passage of a
minimum ionizing particle.  Jets are reconstructed from calorimeter energy
clustered within a cone of radius $\sqrt{
(\Delta \eta )^2 + (\Delta \phi )^2} = 0.5$\cite{coords}.

  We select dilepton top candidate events according to criteria similar to 
those used in our
cross section measurement \cite{top_xs_PRL}.  This selection requires two
leptons, with the transverse energy $E_T$ of each lepton $> 15 (20) $~GeV
for the
$e\mu$ and $\mu\mu$ ($ee$) channels, with $|\eta_e| < 2.5$ and
$|\eta_\mu| < 1.7$, and two or more jets with $E_T > 20$ GeV and 
$|\eta| < 2.5$.
In addition, for the $ee$ and $e\mu$ channels we require significant 
missing transverse energy
$\mbox{${\hbox{$E$\kern-0.6em\lower-.1ex\hbox{/}}}_T$}$ to discriminate
against background sources that have no
final-state neutrinos, while for the 
$\mu\mu$ analysis we reduce the $Z$ boson background by rejecting
 events for which the $\chi^2$ probability of a fit to the
 $Z \rightarrow \mu\mu$ hypothesis is $> 1$\%.
We reject much of the remaining background using the quantities 
$H_T \equiv \sum_{\rm jets}E_T$ and
$H_T^e \equiv H_T + E_T(e_1)$, where all jets with 
$E_T > 15$ GeV and $|\eta| < 2.5$ enter the sum, and $e_1$ is the leading 
$E_T$ electron.  The selection requires $H_T^e (H_T) > 120 (100)$
GeV for the $ee$ and $e\mu$ ($\mu\mu$) channels.  The $ee$ selection reported 
in Ref. \cite{top_xs_PRL} is extended to include an event that contains, in
addition to an electron, one EM energy cluster without an
associated reconstructed track, but with hits in the layers of
the central
tracking system between the interaction vertex and the cluster.  This event
also has a muon near one of its jets, which is evidence of $b$ quark decay, and
further enhances the probability that the event is an example of
$t\overline{t}$ production.  The signal to background ratio for such $b$-tagged 
events in which only one of the EM clusters has an associated track is 
$\approx$ 15/1.  The final sample therefore consists of
three $e\mu$ events, two $ee$ events, and one $\mu\mu$ event, with expected
backgrounds of $0.21 \pm 0.16$, $0.47 \pm 0.09$, and $0.73 \pm 0.25$ events,
respectively (background sources are detailed in  Ref\cite{top_xs_PRL}).  Kinematic
 details for the observed events
can be found in Ref. \cite{varnes_thesis}.
 
    We reconstruct the events according to the $t\overline{t}$ dilepton decay
hypothesis.  After applying the
invariant mass constraints
$m(\ell_1 \nu_1) = m(\ell_2 \nu_2) = m_W$ and
$m(\ell_1 \nu_1 b_1) = m(\ell_2 \nu_2 b_2)$, the system remains  
underconstrained due to the two undetected neutrinos.  We
supply the needed additional constraints by {\sl assuming} values
 for $m_t$ and for two quantities associated with the neutrinos
(as discussed below).
  We then solve for the neutrino momenta up to a fourfold
ambiguity and
assign a weight to each solution to characterize how likely it is to occur in
$t\overline{t}$ production for the assumed $m_t$ \cite{Kondo_paper}.  
We compute the relative weight as a function of assumed $m_t$ for
$80 < m_t < 280$ GeV/{\it c}$^2$, employing two weighting schemes with
differing sensitivities to top production kinematics, decay distributions,
and phase space volume consistent with the event topology.

  The first weighting scheme is called matrix element weighting
(${\cal M}$WT), and is an 
extension of the procedure given in Ref. \cite{DG_paper}.
Here, we require the sum of the neutrino 
$\vec{p}_T$'s to
equal the measured \mbox{$\vec{\hbox{$E$\kern-0.6em\lower-.1ex\hbox{/}}}_T$}.
We assign a weight
\begin{displaymath}
W_o(m_t) = A(m_t)f(x)f(\overline{x})
        P(E_{\ell 1}^{CM} | m_t) P(E_{\ell 2}^{CM} | m_t)
\end{displaymath}
where $f(x)$ is the CTEQ3M \cite{CTEQ3M} parton distribution function 
evaluated at the
proton (antiproton) momentum fraction $x$ ($\overline{x}$) 
required by the solution, and $P(E_{\ell}^{CM}| m_t)$ is the probability
density for the lepton energy in the top
quark rest frame.  The factor $A(m_t)$ normalizes the average weight of
accepted events to unity, independent of the top quark mass.  For 
each mass, we add the weights of all solutions.
            
   The other weighting scheme, called neutrino weighting ($\nu$WT)
 \cite{varnes_thesis}, steps
the assumed $\eta$ for each neutrino through a range of values 
 at each $m_t$.  Each step spans an equal fraction of the 
neutrino $\eta$ distribution expected in  $t\overline{t}$ production. At each
step a weight is
assigned based on the extent to which the
\mbox{$\vec{\hbox{$E$\kern-0.6em\lower-.1ex\hbox{/}}}_T$}
measured in the event agrees with the sum of the neutrino $\vec{p}_T$'s
in the solution. The Gaussian resolution of each
component of the \mbox{$\vec{\hbox{$E$\kern-0.6em\lower-.1ex\hbox{/}}}_T$} is 
4 GeV.  The weights
at all $\eta$ values are summed to give $W_o(m_t)$ in this method.

    In calculating the consistency of the event kinematics with any given
$m_t$, we also account for detector resolution for jets, leptons, and 
$\mbox{${\hbox{$E$\kern-0.6em\lower-.1ex\hbox{/}}}_T$}$.
This is done by fluctuating all the measured energies by their resolutions 
100 (5000) times for MC (collider data) events and summing the weights 
obtained for each fluctuation.  The Gaussian resolutions for electrons and 
muons are  
$\sigma_E/E = 15\%/\sqrt{E} \oplus 3\%$ and
${\sigma (1/p)} = 0.18(p-2 {\rm})/p^2 \oplus  0.003$, respectively, 
with $E$ $(p)$ in GeV (GeV/{\it c}).
  Jets are smeared by double Gaussians
designed to model both the inherent energy resolution of the hadronic 
calorimeter 
(narrow Gaussian) and the contribution of large angle gluon radiation to the
resolution 
(wide Gaussian).   The 
\mbox{$\vec{\hbox{$E$\kern-0.6em\lower-.1ex\hbox{/}}}_T$} is then recomputed to
reflect the changes in jet and lepton energies, and  each component is
fluctuated with a 4 GeV Gaussian.

\begin{figure}[t]
\centerline{\psfig{figure=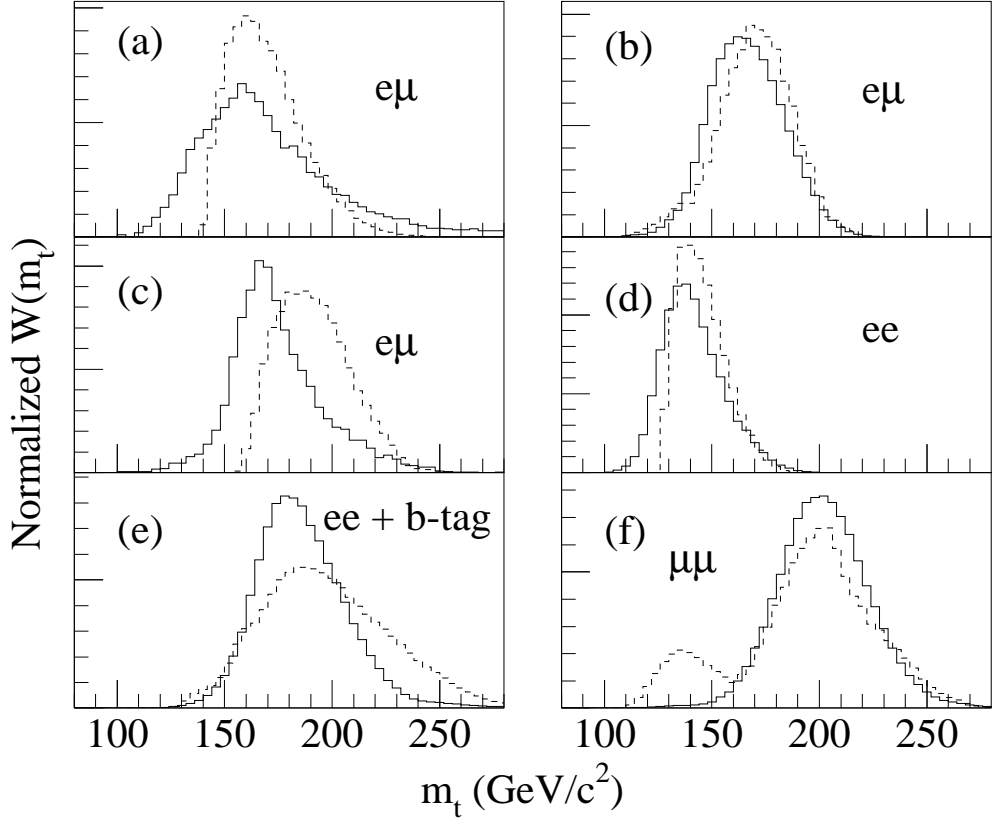,width=6in}}
    \caption{$W(m_t)$ distributions for the six dilepton candidates, using the
${\cal M}$WT (dashed) and $\nu$WT (solid) methods.
\label{fig:wmt}}
\end{figure}

    For both the ${\cal M}$WT and $\nu$WT methods, the up to fourfold solution
ambiguity is
handled by summing the weights for all solutions.
There remains a twofold ambiguity in the pairing of jets to leptons in
 reconstructing the event.  Both analyses sum over both pairings.
  In addition, initial- and final-state gluon radiation (ISR and FSR) 
can create extra jets that further 
complicate the final state.  For events with more than two jets, a weighted 
sum is taken over  all possible combinations of the three leading $E_T$ jets.
If jet $i$ is assumed to arise from ISR, then 
the weight is given by $\exp [-E_{T} \sin \theta_i / (25 \hbox{ GeV})]$ 
\cite{coords}.  
If jets $i$ and $j$ are
assumed to arise from the same $b$ quark by FSR and  have invariant mass 
$m_{ij}$, the weight assigned is
$\exp [-m_{ij} / (20 \hbox{ GeV}/c^2)]$.  In
each case, the form of the weight is based on the characteristics of gluon
radiation and the coefficient is determined empirically to maximize the
sensitivity.  The 
distributions of $W(m_t)$ (which corresponds to $W_o(m_t)$ after accounting 
for resolution and
jet combinations) for the six candidate events, using the two weighting 
methods, are displayed in Fig. \ref{fig:wmt}.

    Due to combinatorics, and the fact that no account has been
taken of background, the $W(m_t)$
distribution derived from either the ${\cal M}$WT or $\nu$WT 
method cannot be considered to be a probability density.  Consequently, a 
maximum likelihood 
fit is performed in which the $W(m_t)$ distributions for data are
compared to the expectations from signal and background.
  The signal is modeled using {\sc herwig} \cite{herwig_paper}
 while the various background sources are modeled by 
{\sc isajet} \cite{isajet_paper}, {\sc pythia} \cite{pythia_paper}, 
and D\O\ data (for the instrumental backgrounds)~\cite{top_xs_PRL}.

    For both analyses, the maximum likelihood fit proceeds by normalizing 
the $W(m_t)$ 
distribution for each event to unity, and
 integrating the fractional
weights in
five 40 GeV/$c^2$ bins in $m_t$.  The first four bins form the
components of a four-dimensional vector $\vec{w}_i$ for each
event $i$.   Using the shape of the $W(m_t)$ distributions
increases the statistical sensitivity of the measurement by $\approx$ 25\%
over a fit to a single-valued mass estimator for each event.
 The 
likelihood $L(m_t,n_s,n_b)$ to be maximized is
\begin{displaymath}
L = g(n_b)p(n_s+n_b) \times \prod_i^N {n_s f_s(\vec{w}_i|m_t) + n_b
 f_b(\vec{w}_i) \over n_s + n_b }\mbox{,}
\end{displaymath}
where $n_s$ and $n_b$ are the fitted signal and background levels,
$g(n_b)$ is a Gaussian constraint that $n_b$
be consistent
with expectations, $p(n_s+n_b)$ is a
Poisson constraint that $n_s+n_b$ be consistent
 with the 
sample size $N$, and $f_s$ and $f_b$ are the four-dimensional 
probability densities for 
signal and background.  The probability densities $f(\vec{w})$ are estimated 
by summing the contributions of four-dimensional Gaussian 
kernels placed at the location of $\vec{w}$ for each event in the signal MC or
background samples~\cite
{PDE_paper}. Using the
 estimated $f_s$ and $f_b$, $L$ is maximized with respect to
$n_s$ and 
$n_b$ for each value of $m_t$ at which we have generated MC events. The 
maximum likelihood estimate of $m_t$
($\widehat m_t$) and its error ($\hat \sigma$) are determined by a quadratic
 fit to $-\ln L$ for the nine points about the minimum.
    
\begin{figure}[t]
\centerline{\psfig{figure=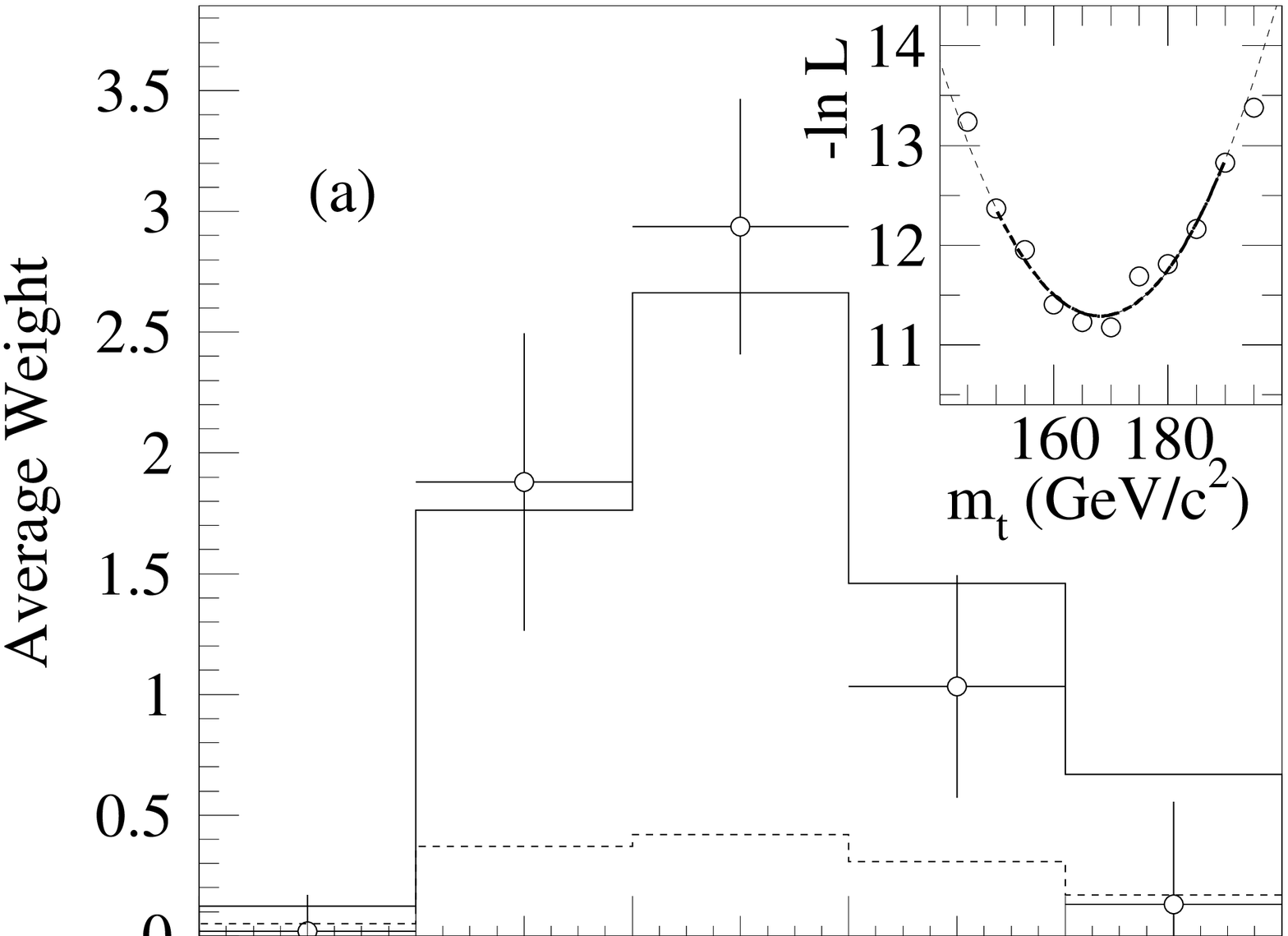,width=3.8in}}
\centerline{\psfig{figure=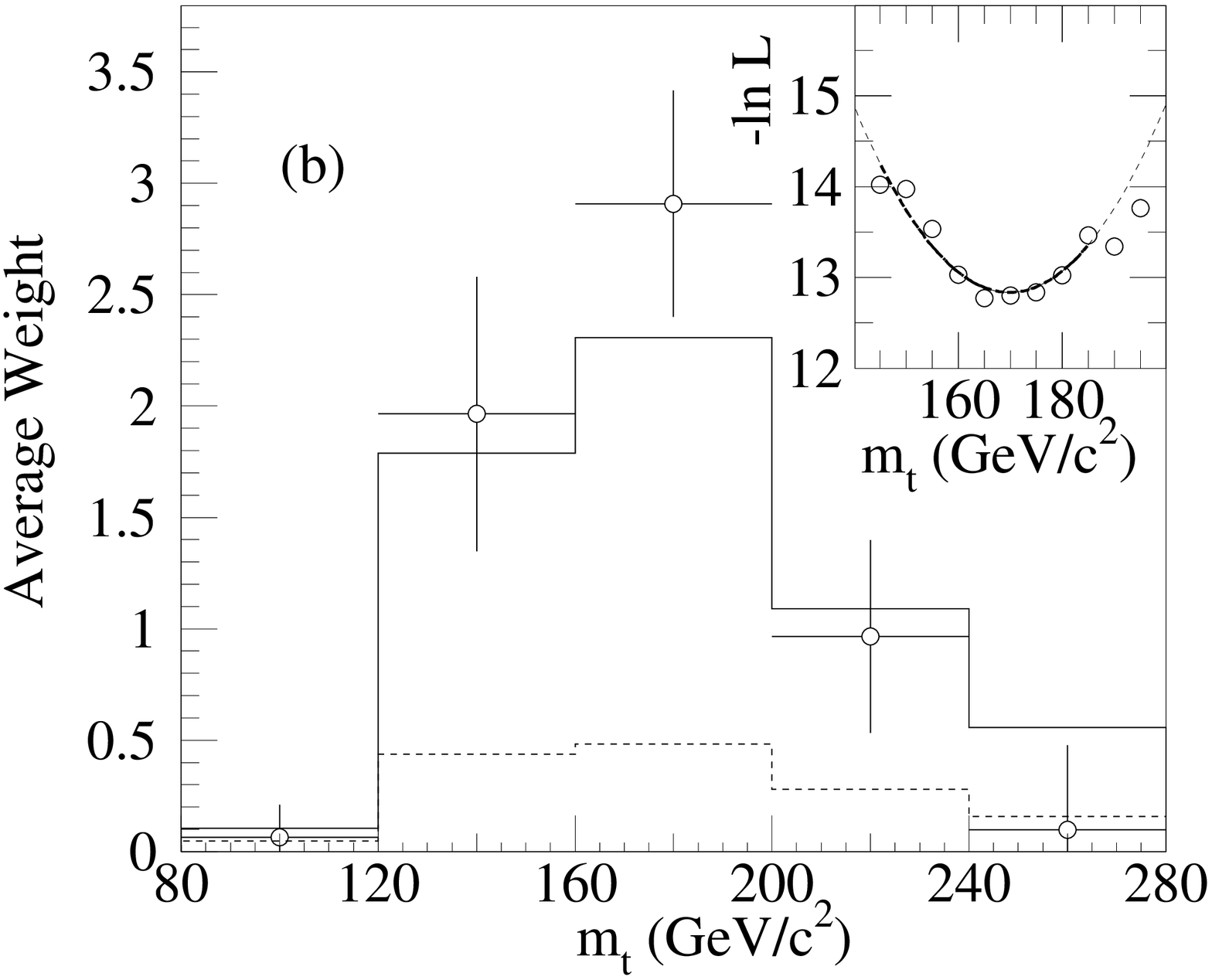,width=3.8in}}
    \caption{Sum of the normalized candidate weights grouped into the five
bins considered in the maximum likelihood fit (circles) for the (a) 
${\cal M}$WT and (b) ${\nu}$WT analyses.  The uncertainty on these points is
taken from the RMS spread of the weights in MC studies.  Also shown are the
average weights from the best-fit background (dashed) and signal plus
background (solid).  The $-\ln L$ distributions and quadratic fits are 
inset.
   \label{fig:dilep_results}}
\end{figure}

    Applying the maximum likelihood fit to the data, we determine
 the top quark mass  to be
$m_t = 168.1 \pm 12.4$ GeV/{\it c}$^2$ (${\cal M}$WT), and
$m_t = 169.9 \pm 14.8$ GeV/{\it c}$^2$ ($\nu$WT), where the uncertainties are
statistical only (see Fig. \ref{fig:dilep_results}).  The results of fits to 
subsamples of the data are listed in
Table~\ref{tab:results}.

\begin{table}
\caption{Summary of $m_t$ measurements for full and partial data sets.
Uncertainties are statistical only. 
\label{tab:results}}
\begin{tabular}{lcc} 
 Channels Fit  &  ${\cal M}$WT & $\nu$WT \\ \hline
$e\mu$ + $ee$ + $\mu\mu$ 
 &  $168.1 \pm 12.4$ GeV/{\it c}$^2$ & $169.9 \pm 14.8$ GeV/{\it c}$^2$  \\
$e\mu$ + $ee$             
 & $167.9 \pm 12.6$ GeV/{\it c}$^2$  & $173.2 \pm 14.0$ GeV/{\it c}$^2$\\
$e\mu$                   
 & $173.1 \pm 13.3$ GeV/{\it c}$^2$  & $170.1 \pm 14.5$ GeV/{\it c}$^2$\\
\end{tabular}
\end{table}

\begin{figure}
\centerline{\psfig{figure=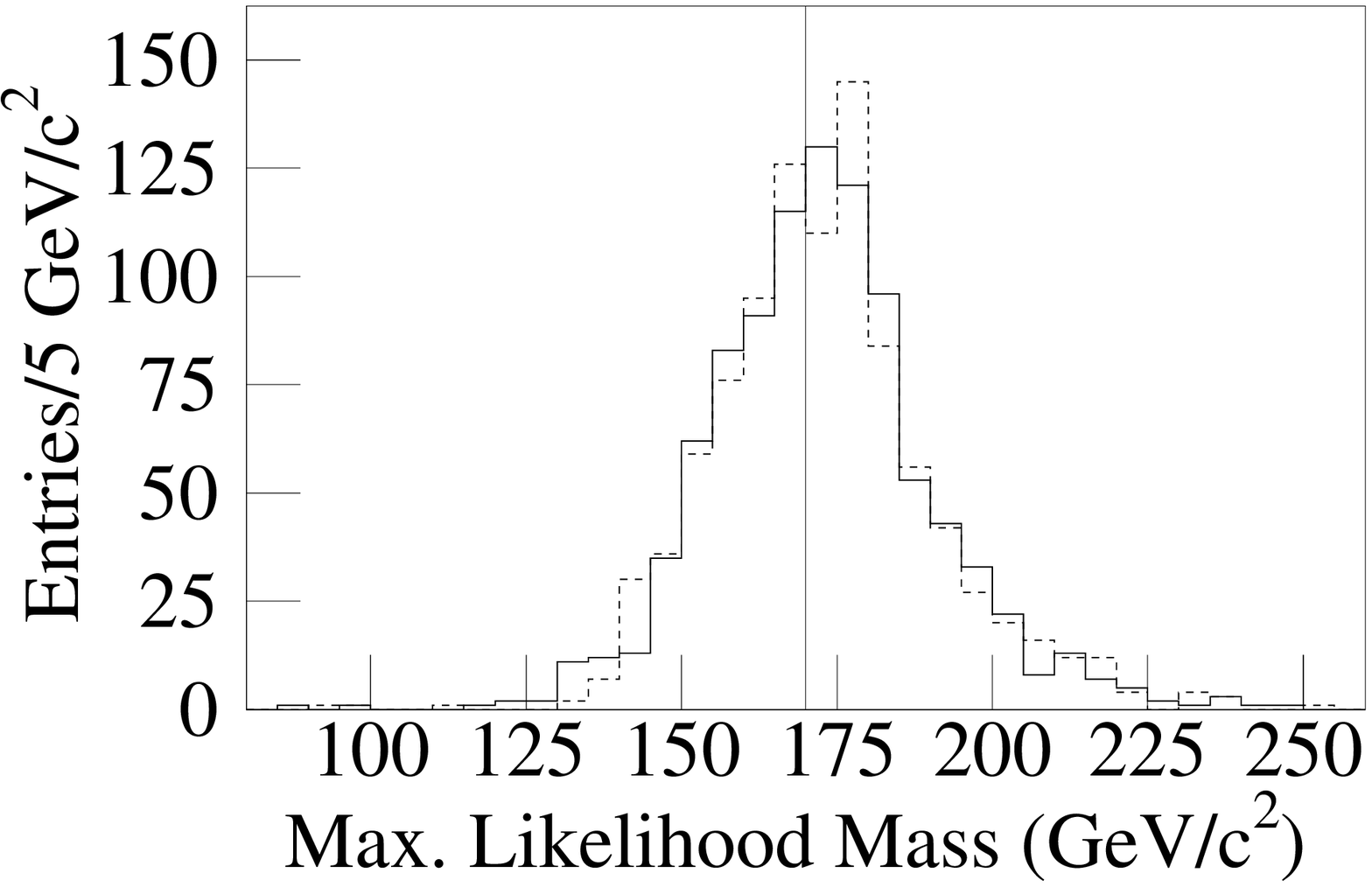,width=4in}}
\caption{Distribution of $\widehat m_t$ for ``experiments'' with
 $m_t^{MC}=170$ GeV/{\it c}$^2$ (marked by vertical line)
using the ${\cal M}$WT (dashed) and $\nu$WT (solid) analysis methods.
\label{fig:ensems}}
\end{figure}

  To study the properties of our mass estimator, 1000 MC ``experiments'' are 
generated
by randomly selecting three $e\mu$, two $ee$, and one $\mu\mu$
events from simulated top (with input mass $m_t^{MC}$) and background samples,
according to the estimated
background contamination. Due to the small sample size the estimator is not
normally distributed. Figure~\ref{fig:ensems} shows a typical distribution
for $\widehat m_t$.  The distribution is characterized by its median and width 
(half of the shortest interval that contains 68.3\% of the 
experiments).  The width of a Gaussian fit to the pull distribution
($\widehat m_t-m_t^{MC})/ \hat \sigma$ is consistent with unity.  This verifies 
that $\hat \sigma$ is an unbiased estimate of the statistical uncertainty.
The pulls means can
differ from zero because of the non-Gaussian tails of the $\widehat m_t$
distribution. 
Table II lists medians, widths, and means and widths of the pulls for different
$m_t^{MC}$. The properties of $\widehat m_t$ are very 
similar for both methods.

\begin{table}[b]
\caption{Properties of the maximum likelihood estimate $\widehat m_t$.  
 \label{tab:ensem_tests}}
\begin{tabular}{ddddcd} 
  &  $m_t^{MC}$  &  Median &  Width & Pull & Pull  \\
  &  GeV/{\it c}$^2$   &   GeV/{\it c}$^2$  &   GeV/{\it c}$^2$  
  & Mean  & Width \\  \hline
        & 160   &  161.6    &  15.8  & 0.12 $\pm$ 0.03 & 1.03 \\
${\cal M}$WT & 170    & 172.2     &  16.7  & 0.11 $\pm$ 0.03 & 0.99  \\
        & 180   &   180.5     &  17.3  & 0.00 $\pm$ 0.03 & 0.98  \\ \hline
        & 160    &  161.5   &  14.4  & 0.17 $\pm$ 0.03 & 0.96  \\
$\nu$WT & 170    &  172.2   &  16.2  & 0.08 $\pm$ 0.03 & 0.98  \\
        & 180    &  180.5   &  18.1  &  0.03 $\pm$ 0.03 & 1.03  \\ 
\end{tabular}
\end{table}

    The sources of systematic uncertainty are summarized in Table
\ref{tab:syst}.  A major
uncertainty arises from the jet energy calibration, which proceeds in two
steps.  The first step uses $\gamma +$jets events to relate the
well-calibrated electromagnetic scale to the hadronic scale.  Effects 
considered include the overall hadronic response, energy added to jets by
multiple interactions and uranium noise, and the spread of the hadronic 
shower outside of the jet cone \cite{bob_talk}.  The second step follows from
a detailed comparison of  $\gamma +$ jets events in MC and collider data,  
providing a
correction that depends on jet $\eta$ and ensures
that the energy scale in data matches that in MC.  An additional correction 
is applied to jets having a muon within the jet cone. 
Under the assumption that the
muon was produced in the semileptonic decay of a heavy quark, the jet 
energy is increased take into account the muon and neutrino 
energies.  
After these corrections, we find (using the $\gamma+$jets and smaller
$Z(\rightarrow ee)$ + jets samples) that the data and MC scales agree,
with an uncertainty of  $\delta(E_T) = 0.025 E_T + 0.5 \mbox{ GeV}$.  
MC tests are run on
samples with the jet energies rescaled by $\pm \delta$ to yield an energy scale
uncertainty of 2.4 GeV/$c^2$ in $m_t$.

\begin{table}
\caption{Systematic errors in the measurement of $m_t$.
\label{tab:syst}}
\begin{tabular}{lc} 
 Source   & Error (GeV/{\it c}$^2$)  \\ \hline
Jet energy scale  &   2.4\\
Signal model          &  1.8 \\
Multiple interactions  & 1.3  \\
Background model      &  1.2  \\
Likelihood fit        &  1.3 \\ \hline
Total                 &  3.7  \\
\end{tabular}
\end{table}

    Other sources of uncertainty arise from differences among 
models of $t\overline{t}$
and background production, multiple interactions, and the likelihood fit
procedure.
{\sc isajet} is
used as a cross-check of the {\sc herwig} $t\overline{t}$ production
model.
The effect of multiple interactions is 
estimated using MC samples with additional random $p\overline{p}$ interactions 
overlaid
on $t\overline{t}$ production.  The contributions from all
sources are summed in quadrature to give the systematic uncertainty on the
measurement (see Table \ref{tab:syst}).
    
    Taking account of the 77\% correlation between the ${\cal M}$WT and $\nu$WT
analyses, we measure the mass of the top quark in the dilepton channel by
combining the two results: 
\begin{displaymath}
m_t = 168.4 \pm 12.3 \mbox{ (stat)} \pm 3.7 \mbox{ (sys) GeV/{\it c}$^2$}
\mbox{.} 
\end{displaymath}
This value is in good agreement with the measurement from the
 $\ell$ + jets channel \cite{ljets_mass_PRL}, consistent with the
 $t\overline{t}$ hypothesis for both channels.  We combine the results of the
single- and dilepton analyses by propagating the systematic uncertainties in 
each channel with
correlation coefficients of either zero
(for MC statistics and background model) or 
unity (for jet energy scale, multiple interactions, and $t\overline{t}$ 
production models).  Using 
all dilepton and $\ell$ + jets 
$t\overline{t}$ candidates, we measure:

\begin{displaymath}
m_t = 172.0 \pm 5.1 \mbox{ (stat)} \pm 5.5 \mbox{ (sys) GeV/{\it c}$^2$}
\mbox{,}
\end{displaymath}
or, combining statistical and systematic errors in quadrature, 
$m_t = 172.0 \pm 7.5$ GeV/c$^2$.

%
We thank the staffs at Fermilab and collaborating institutions for their
contributions to this work, and acknowledge support from the 
Department of Energy and National Science Foundation (U.S.),  
Commissariat  \` a L'Energie Atomique (France), 
State Committee for Science and Technology and Ministry for Atomic 
   Energy (Russia),
CNPq (Brazil),
Departments of Atomic Energy and Science and Education (India),
Colciencias (Colombia),
CONACyT (Mexico),
Ministry of Education and KOSEF (Korea),
CONICET and UBACyT (Argentina),
and the A.P. Sloan Foundation.


\begin{references}
\bibitem{coords}
D\O\ uses a cylindrical coordinate system about the beam axis, with polar
angle $\theta$ and azimuthal angle $\phi$.  The pseudorapidity  
$\eta \equiv \ln \left( \cot \theta /2 \right)$ is also used.
\end{references}
\end{document}